\documentclass{aastex}          
\usepackage{spr-astr-addons}    
\begin{document}

\title{Five steps in the evolution from protoplanetary to debris disk}

\shorttitle{From protoplanetary to debris disk}
\shortauthors{Wyatt et al.}

\author{M. C. Wyatt} 
\and 
\author{O. Pani\'{c}}
\and 
\author{G. M. Kennedy}
\and 
\author{L. Matr\`{a}}
\affil{Institute of Astronomy, University of Cambridge, Madingley Road, Cambridge CB3 0HA, UK}
\email{wyatt@ast.cam.ac.uk} 



\begin{abstract}
The protoplanetary disks seen around Herbig Ae stars eventually dissipate leaving
just a tenuous debris disk, comprised of planetesimals and the dust derived from them,
as well as possibly gas and planets.
This paper uses the properties of the youngest (10-20\,Myr) A star debris disks to consider
the transition from protoplanetary to debris disk.
It is argued that the physical distinction between these two classes should rest on the presence
of primordial gas in sufficient quantities to dominate the motion of small dust grains
(rather than on the secondary nature of the dust or its level of stirring).
This motivates an observational classification based on the dust emission spectrum which
is empirically defined so that A star debris disks require fractional excesses $<3$
at 12\,$\mu$m and $<2000$ at 70\,$\mu$m.
We also propose that a useful hypothesis to test is that the planet and planetesimal systems seen on
the main sequence are already in place during the protoplanetary disk phase, but are
obscured or overwhelmed by the rest of the disk.
This may be only weakly true if the architecture of the planetary system continues to change 
until frozen at the epoch of disk dispersal,
or completely false if planets and planetesimals form during
the relatively short dispersal phase.
Five steps in the transition are discussed:
\textbf{(i)} the well-known carving of an inner hole to form a \textit{transition disk};
\textbf{(ii)} depletion of mm-sized dust in the outer disk, where it is noted that it is
of critical importance to ascertain whether this mass ends up in larger planetesimals or is
collisionally depleted;
\textbf{(iii)} final clearing of inner regions, where it is noted that multiple debris-like
mechanisms exist to replenish moderate levels of hot dust at later phases, and that these likely
also operate in protoplanetary disks;
\textbf{(iv)} disappearence of the gas, noting the recent discoveries of both primordial and
secondary gas in debris disks which highlight our ignorance in this area and its impending
enlightenment by ALMA;
\textbf{(v)} formation of ring-like structure of planetesimals, noting that these are shaped by interactions
with planets, and that the location of the planetesimals in protoplanetary disks may be
unrelated to that of dust concentrations therein that are set by gas interactions.
\end{abstract}




\section{Introduction}
\label{s:intro}
While there have been many advances in our understanding
of the structure of protoplanetary disks, there remain considerable
unknowns, in particular with respect to the status of planet formation
within them, and the processes that ultimately make these disks
disappear.
Protoplanetary disk evolution on Myr timescales is mainly driven by accretion, though
photoevaporation may also drive significant mass loss
\citep[see][]{Hollenbach1994}, and is especially dominant in the presence of
external sources of radiation \citep[e.g., in Orion,][]{Mann2014}.
These dispersal processes directly affect the amount of material available for planet
formation, and so presumably the eventual architecture of planetary systems, although
planet formation processes may themselves also contribute to shaping a disk's structure
and evolution, especially in the case that massive planets have formed.

In the innermost regions of protoplanetary disks, gap and hole opening is believed to be the first
detectable step away from a continuous primordial disc and towards a more radially
concentrated structure, and can be initiated by photoevaporation or planet formation.
The first evidence of this inside-out evolution was seen in disks lacking
near-infrared excesses \citep[e.g.,][]{Strom1989, Skrutskie1990} which were consequently
named transition disks.
However, the subsequent steps in that evolution are very poorly constrained,
especially the relative timescales on which gas and dust dissapear
(see, e.g., \S \ref{ss:gas}).

We know that the end-state of the evolution is for the star to become a main sequence
star with (possibly) a remnant debris disk along with a planetary system.
Debris disks are discovered like their protoplanetary counterparts by an excess of
infrared emission above that expected from the star itself, and because
most nearby stars are on the main sequence, their disks can be imaged in
great detail and dust detected down to very low levels \citep{Wyatt2008,Eiroa2013}.
The physical picture of debris disks is that the dust we see is continually
replenished by collisions between planetesimals in orbit around the star.
Thus they are considered to be analogous to the asteroid and Kuiper belts in
the Solar System, and as such are usually viewed as being a component of the star's
planetary system that formed during the protoplanetary disk phase.
This idea is reinforced in several systems in which planets are known
in the regions interior to debris disks \citep{Marois2008,Kalas2008,Marshall2014,Rameau2013}.

In this paper we will consider debris disks as descendants of protoplanetary
disks and use that perspective to piece together the evolutionary stages that
protoplanetary disks must go through after the transition disk phase.
In particular we will focus on the properties of the disks at the stage immediately
after the protoplanetary disk has dispersed.
Section~\ref{s:diff} considers the difference between protoplanetary and debris
disks, both from an observational and physical point of view.
Section~\ref{s:stages} then outlines five steps that must occur in
the evolution, focussing either on different radial locations, or the
evolution of different material components of the disks.

As there are aspects of the evolution that may depend on the mass of the
star, either because of how the stellar luminosity affects the disk's evolution
\citep[e.g.,][]{KennedyKenyon2009}, or
because of how the stellar properties affect the detectability of the disk
\citep{Wyatt2008}, here we will focus the discussion on the evolution of
intermediate mass stars.
That is, we will consider the evolution of Herbig Ae (hereafter HAe) stars
into main sequence A stars.
This choice benefits from A star debris disks being more readily detectable than
those around their lower stellar mass counterparts, resulting in better constraints
on the structure of their disks around the transition stage.
Also, a lot of observational effort in understanding the latter stages of protoplanetary disk
evolution has focused on HAe stars, which are generaly brighter than their lower mass
T Tauri star counterparts.
However, a downside is that HAes are rare and diverse in their ages and formation
environment, and they therefore do not form an unbiased sample.

\section{Difference between protoplanetary and debris disks}
\label{s:diff}
There is no formal distinction between protoplanetary and debris disks,
as will become clear in this section.
A crude categorisation can be achieved based solely on the age of the host star,
with the division at roughly 10\,Myr.
However, there are counter-examples on either side of this division;
e.g., there are examples of 30\,Myr-old protoplanetary disks \citep{Scicluna2014},
and of debris disks in young regions like Upper Sco
\citep{Carpenter2009} and the TWHya Association \citep{RiviereMarichalar2013}.
Indeed, detection of a debris disk at the distance of most pre-main sequence stars ($\sim 140$\,pc)
is limited by observational capabilities, so it is possible that many of the $<10$\,Myr stars
currently without detectable disks will turn out to have debris disks with deeper observations.
A more fundamental difference is the amount of dust;
debris disks are usually optically thin whereas protoplanetary
disks are optically thick\footnote{In this paper optical depth refers to the fraction of starlight that
reaches disk material directly, and so refers to the depth at optical and UV wavelengths along the
line-of-sight to the star.}, which is also evident as more than an order of
magnitude difference in the inferred mass of mm-sized dust
\citep[][see \S \ref{ss:mm}]{Wyatt2003a,Panic2013}.
The radial distribution of the material is also different, with debris disks
generally being narrower with large inner holes and protoplanetary disks much
broader; 
though again there are counter-examples such as the transition disk class of
protoplanetary disk which are characterised by large inner holes.
Crucially the large mass of gas, predominantly primordial H$_2$, which is the main mass component of
a protoplanetary disk, is no longer present in debris disks.
Even though the gas in a protoplanetary disk is not always readily detectable,
it can still have a strong influence on the structure of the dust disk;
e.g., the gas pressure keeps the dust disk vertically thick causing larger excess emission.

\subsection{Disks around Herbig Ae stars}
\label{ss:ppd}
While statistical studies of HAe stars suffer from a relatively small number of targets
compared to those of lower mass stars, the HAe stars possess brighter disks and consequently
host some of the best studied protoplanetary disks in terms of their detailed structure.
Most of the known HAe stars were first identified as potentially young 
intermediate-mass stars by \citet{Herbig1960} using the criteria of emission line spectra, 
regions of heavy obscuration and illumination of a nearby nebulosity.
\citet{FinkenzellerMundt1984} studied the properties of the stars in the 
\citet{Herbig1960} sample and established that these had significantly higher infrared 
excesses than ordinary Be stars, while
infrared studies by \citet{DongHu1991} and \citet{Berrilli1992} further confirmed
the existence of circumstellar material and so the pre-main sequence nature of these stars,
as suspected from their spectral characteristics.
\citet{Grinin1989, Grinin1991} pointed out the existence of observationally 
similar stars to the HAe group, but in relative isolation from nebulosities,
which were later included in this denomination.
A catalogue summarising this class is given in 
\citet{The1994}, and typical ages for HAes were found to be 5-10\,Myr
\citep{vandenAncker1997}.

These disks are traditionally classified from their infrared spectral slopes into
group I and II sources \citep{Meeus2001}, which are interpreted as flared and flat structures,
respectively, that are linked to the degree of grain growth and settling in the disks
\citep[e.g.,][]{Dullemond2004}.
Millimetre spectral slopes provide further evidence of grain growth in these sources
\citep[e.g.,][]{Acke2004}.
The relatively low level of near-infrared emission in HAe disks suggest that most
have inner regions that are relatively empty of dust
\citep{Yasui2014}, 
an interpretation which is confirmed for the disks that have been directly imaged
in scattered light \citep{Fukagawa2006}, polarised light \citep{Quanz2012a}
or submillimetre emission \citep{Brown2012}.
While group II sources tend to have larger grains than those in group I,
group II disks are not necessarily at a later evolutionary stage, since there is also
a high incidence of transition disks in group I \citep{Maaskant2013}.

Most recently, ALMA has allowed breakthroughs in understanding transitional HAe disks,
for example finding asymmetries in the dust distribution outside the gap
\citep{vanderMarel2013}, large discrepancies between the radial distributions
of gas and dust \citep{Pineda2014, Walsh2014}, and structures
associated with the flow of gas across the dust gap \citep{Casassus2013}.
Many of these features have been linked to putative planets.
For example, studies of planet-disk interactions show that 
planets can open gaps in the gas disk, significantly disturb the inward
flow of gas, and induce pressure waves that result in spiral features
propagating through the disk \citep[e.g.,][]{Paardekooper2007}.
The reaction of dust of different sizes to these gas disk structures explains
many of the observed features \citep[e.g.,][]{Pinilla2012}.

Here we describe a few HAe sources where the disk structure has been studied in great detail,
and which exhibit a range of structural features.
While these sources are not necessarily representative, and in no way form an unbiased
sample, they will be useful for reference later in the paper.

\textbf{HD100546}:
This 8-10\,Myr star \citep{vandenAncker1997} has a disk that still contains 
significant amounts of gas \citep{Panic2010, Thi2013a} and its dust mass
undoubtedly places it in the group of protoplanetary disks (see \S \ref{ss:mm}).
There is a substantial decrease in gas and dust density inside 10\,au seen in
spatially resolved observations \citep{Wilner2003, vanderPlas2009, Quanz2011}
further identifying this as a transition disk,
as also inferred from modelling of the Spectral Energy Distribution \citep[SED;][]{Bouwman2003}.
Dust is detected interferometrically in the near- and mid-infrared at $<0.7$\,au
\citep{Benisty2010, Panic2014}, so rather than an empty inner hole
this disk has a very wide gap.
The gap has often been linked to a putative massive planet at that location
\citep[$60M_{\rm{Jup}}$,][]{Mulders2013}.
\citet{Quanz2013} also report detection of a more distant planet at 69\,au, indirect support for
which comes from millimetre imaging that shows non-co-spatial dust and gas distributions
as expected from interactions with such a planet, with the outer disk ($>60$\,au) depleted
of millimetre dust while still abundant in gas \citep{Pineda2014, Walsh2014}.
Spiral features are seen in scattered light at $\sim$100\,au scales out to 350\,au.
Amongst the numerous observational tracers, strong PAH emission arises from the surface of this disk
at a range of radii.

\textbf{HD141569}: This 5\,Myr old B9.5 star \citep{Weinberger2000, Merin2004}
has an infrared excess with a fractional luminosity of 0.0084 \citep{Sylvester1996}
comparable to the most luminous debris disks.
However, detections of CO \citep{Zuckerman1995, Dent2005} and Polycyclic Aromatic Hydrocarbon
(PAH) emission
\citep{Keller2008, Geers2009} discern this source from debris disks suggesting a younger
evolutionary stage, with large amounts of both gas \citep[$10^{-4}M_\odot$,][]{Thi2014}
and small dust \citep{Merin2004} present.
Indeed, its sub-mm dust mass of $0.7M_\oplus$ \citep{Sandell2011, Panic2013}
and excess emission over a range of wavelengths suggest that this disk is
intermediate between the old evolved protoplanetary disks
and very young and luminous debris disks.
The CO line profile implies a radially broad (90-250\,au) gas distribution
\citep{Dent2005, Thi2014},
with an additional warm component that is detected spectro-astrometrically at 17-50\,au
\citep{Brittain2002}.
There is little information on the inner 100\,au region, as it is masked in imaging at
short wavelengths, though mid-IR images confirm that there is dust in the inner regions
\citep{Fisher2000, Maaskant2014}.
In the outer disk, scattered light imaging \citep{Clampin2003} shows two belts of emission between
175 and 400\,au, with a gap between 215 and 300\,au and several spiral features.
While the open spirals at $>400$\,au may be caused by a flyby of a known stellar companion at
1000\,au projected separation \citep{Quillen2005, Ardila2005}, the more tightly
wound ones at 200 and 325\,au are best explained by internal perturbations from a 
(yet unseen) planet embedded in the
disk \citep{Wyatt2005}.

\begin{table*}
\small
\caption{Properties of the four A stars from the TWA and BPMG with debris disks}
\label{tbl:young}
\begin{tabular}{llll}
\tableline  
Star & Hot Dust & Cold Dust & Gas \\
\tableline  
HR4796A (A0V, 8\,Myr, 67\,pc) & No? & Narrow ring 80\,au $f=5\times10^{-3}$ & No \\
$\beta$ Pic (A5V, 20\,Myr, 19\,pc) & $\sim 10$\,au $f=0.9 \times 10^{-3}$ & Broad ring 60-130\,au $f=1.5 \times 10^{-3}$ & CO (+...) \\
$\eta$ Tel (A0V, 20\,Myr, 48\,pc) & 4\,au $f=1.6 \times 10^{-4}$ & 24\,au $f=1.4 \times 10^{-4}$ & CII \\
HD172555 (A5V, 20\,Myr, 29\,pc) & 1-8\,au $f=0.7\times10^{-4}$ & None detected & OI \\
\tableline 
\end{tabular}
\end{table*}

\subsection{Young (8-20\,Myr) A star debris disk sample}
\label{ss:dd}
The circumstellar environments of main sequence
A stars evolve on timescales of $\sim 150$\,Myr \citep{Rieke2005},
and perhaps even faster \citep{Currie2008}.
Thus to capture the structure of debris disks in the immediate aftermath
of the dispersal of the protoplanetary disk we need to consider the youngest
known debris disks.
To construct an unbiased sample we use the nearby $\sim 20$\,Myr $\beta$ Pic moving group
\citep[BPMG;][]{Zuckerman2001,Binks2014,Mamajek2014}
and the 8\,Myr TWHya association \citep[TWA;][]{Ducourant2014}.
These associations contain 6 main sequence A stars of which
4 are known to host debris disks, the properties of which are
summarised in Table~\ref{tbl:young} and discussed in greater depth below.

\textbf{HR4796A}: The debris ring in this system has been imaged both in mid-IR and
scattered light and found to lie in a narrow ring at 80\,au in radius, with
both a brightness asymmetry and offset ring centre implying that the ring
is eccentric \citep{Moerchen2011,Thalmann2011,Wahhaj2014,Perrin2014}.
The presence of hot dust is debated \citep{Wahhaj2005,RiviereMarichalar2013,Chen2014,Kennedy2014b}.
No CO was detected suggesting an upper limit of $7M_\oplus$
on the mass of molecular H$_2$ gas present \citep{Zuckerman1995,Greaves2000a},
with more recent upper limits on OI from Herschel \citep{Meeus2012}.

\textbf{$\beta$ Pic}: Scattered light from this edge-on debris disk is seen out to
1000s of au \citep{Smith1984}, but the planetesimals creating the dust
are concentrated between 60-130\,au \citep{Augereau2001,Dent2014}.
The structure of the disk includes many asymmetries including a warp at 80\,au
which is at a similar spatial scale to a clump seen in both $\mu$m-sized
dust and CO \citep{Telesco2005,Dent2014}.
The inner regions of the system are not completely empty, hosting both a giant
planet orbiting at 9\,au \citep{Lagrange2010}, hot dust seen in the
emission spectrum inferred to lie at a comparable distance \citep[e.g.,][]{Okamoto2004},
as well as transient absorption features along the line-of-sight to the star inferred to
originate in distintegrating planetesimals \citep[so-called Falling Evaporating Bodies,
or FEBs;][]{VidalMadjar1994}.
Resolved imaging of metallic ions \citep{Brandeker2004,Nilsson2012} and
of the CO emission \citep{Dent2014} shows the spatial distribution of gas
in this system, further constraints on which are found from far-IR CII lines
\citep{Cataldi2014}.
The gas is located in the outer disk and is thought to originate in the planetesimals.

\textbf{HD172555}: Hot dust was detected in the mid-IR spectrum of this star 
from which the composition of the dust could also be determined
\citep{Chen2006,Lisse2009}.
This hot emission was marginally resolved placing it between 1-8\,au
\citep{Moerchen2010,Smith2012a}.
The absence of far-IR emission above that expected from the hot dust
suggests there is no cold dust in this system \citep{RiviereMarichalar2014}.
The presence of SiO gas was suggested from the mid-IR spectrum \citep{Lisse2009,Johnson2012},
and OI emission was detected in the far-IR implying a mass of gaseous oxygen
$>10^{-4}M_\oplus$ \citep{RiviereMarichalar2012};
variable CaII absorption features along the line-of-sight to the star also
suggest the presence of FEBs \citep{Kiefer2014}.
Another example of a young hot dust-only A star is $\sim 10$\,Myr-old A9V EF Cha
\citep{Rhee2007}.

\textbf{$\eta$ Tel}: The debris disk of this system was resolved in the mid-IR at
a radius of 24\,au, but it was shown that there is an additional
unresolved component at 4\,au \citep{Smith2009}.
The fractional luminosities of the two components
determined from a two temperature fit to the SED show these are of comparable
luminosity.
Gaseous CII (but no OI) was found in the far-IR spectrum giving a mass of gaseous
carbon of $>1.6 \times 10^{-4}M_\oplus$ \citep{RiviereMarichalar2014}.
The star has an M7/8V binary companion at 192\,au projected separation
\citep{Lowrance2000}.
Other young A stars with CII but no OI include 40\,Myr-old 49 Ceti \citep{Zuckerman2012,Roberge2013} and 30\,Myr-old HD32297 \citep{Donaldson2013}.
Also notable are $\sim 10$\,Myr-old A stars HD131488 and HD121191 that both have
two component disks \citep{Melis2013}.

With such small numbers there is no guarantee that the outcomes in the above
sample are representative.
Furthermore, stars in this sample are all members of nearby moving groups, rather than
say members of larger clusters or stars formed in isolation, which is another
reason that they may not represent an unbiased sample of the progenitors of
older main sequence A stars.
Nevertheless, the above sample gives a sense of the diversity in location, width
and composition of young debris disks, as well as the fraction of HAe disks
that might end up in different outcomes.
Another young disk close to this age, but not part of the unbiased sample,
that we will also use to highlight one aspect of the evolution is HD21997.

\textbf{HD21997}: This A3IV/V member of the 30\,Myr Columba association
rose to prominence because of the high fractional luminosity of
its debris disk \citep[$f=5 \times 10^{-4}$;][]{Moor2006},
and the subsequent detection of CO \citep{Moor2011} and imaging of its disk
with ALMA \citep{Moor2013,Kospal2013}.
Another main sequence A star of comparable age with CO is 40\,Myr-old 49 Ceti
\citep{Dent2005,Hughes2008}.

\subsection{Observational classification}
\label{ss:obs}
As mentioned above, there is no accepted observational definition of
a debris disk.
For some authors the distinction lies in the total fractional luminosity
(the ratio of the infrared luminosity to that of the star, $f=L_{\rm{IR}}/L_\star$),
since this approaches 1 for the brightest HAe disks
(e.g., $f \approx 0.5$ for HD100546), with slightly lower values $f<0.25$ for group II 
HAe disks \citep{KenyonHartmann1987}, and the HAe stars closest to the
transition towards debris disks have $f \approx 0.01$
(e.g., $f=0.01$ for HD141569 and $f=0.03$ for 51 Oph).
Thus the dividing line is usually placed at $f \approx 0.01$, which is around the 
level at which the dust might be expected to become optically thin in the radial direction
to the starlight (i.e., in the optical and UV).
However, the transition from optically thin to optically thick depends on
the degree of settling in the disk, and it may be possible to
make a protoplanetary disk model with a lower fractional luminosity as
long as it has a low dust scale height.
It may, however, be hard to maintain such
a low height for small dust if it is entrained in the gas which inevitably has a vertically
broad distribution due to its pressure support.
It is also possible to make a debris disk model with a higher level of
fractional luminosity, since the disruption of a single large asteroid, if 
in close proximity to the star, can release sufficient
cross-sectional area to intercept a large fraction of the starlight.
These issues were discussed in \citet{Kennedy2014} in relation to the
disk of HD166191 for which different interpretations are invoked by
different authors, highlighting the difficulty of coming up with an
independent classification scheme.

\begin{figure*}
\begin{tabular}{cc}
\includegraphics[width=3.1in]{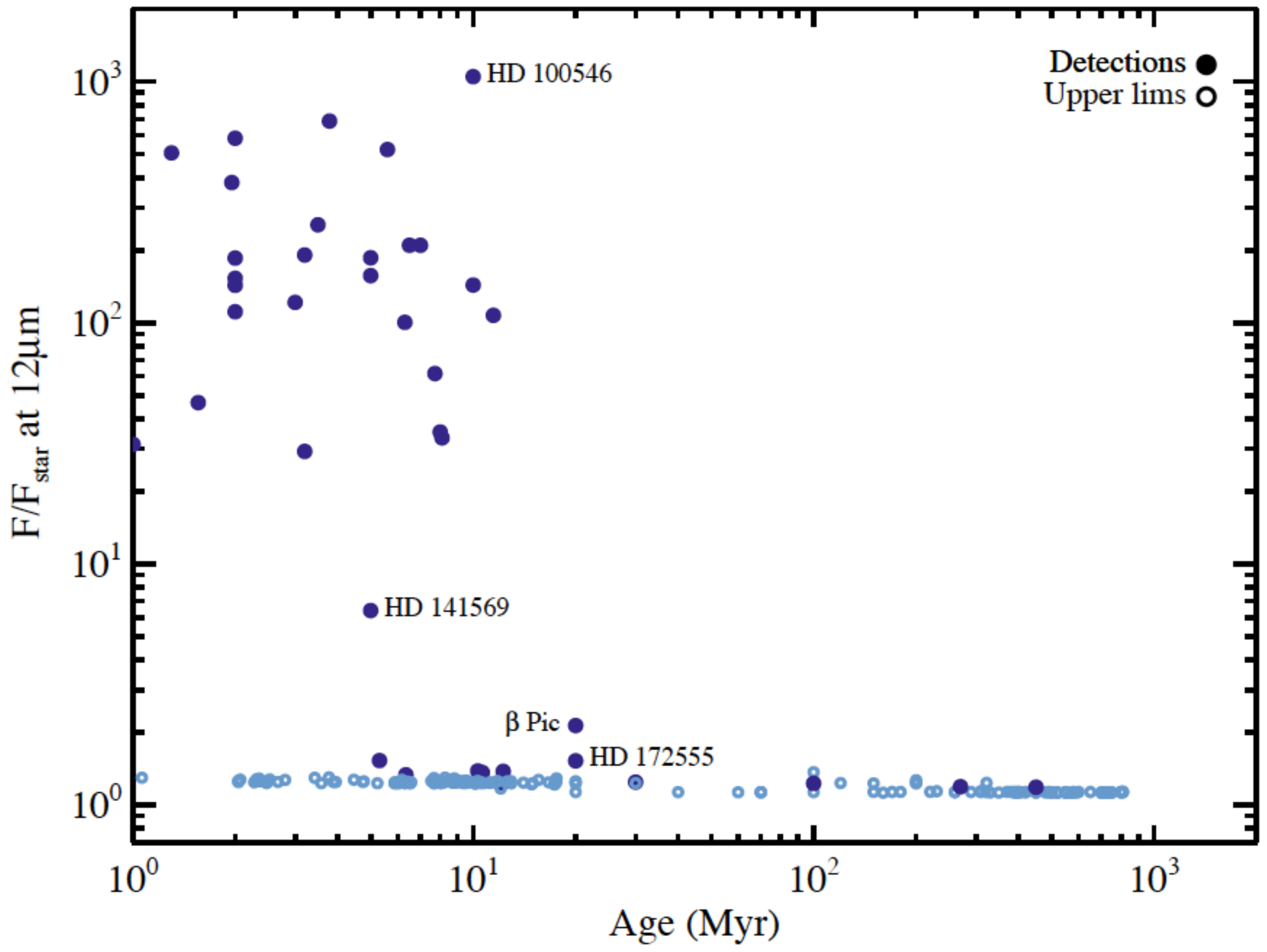} &
\includegraphics[width=3.1in]{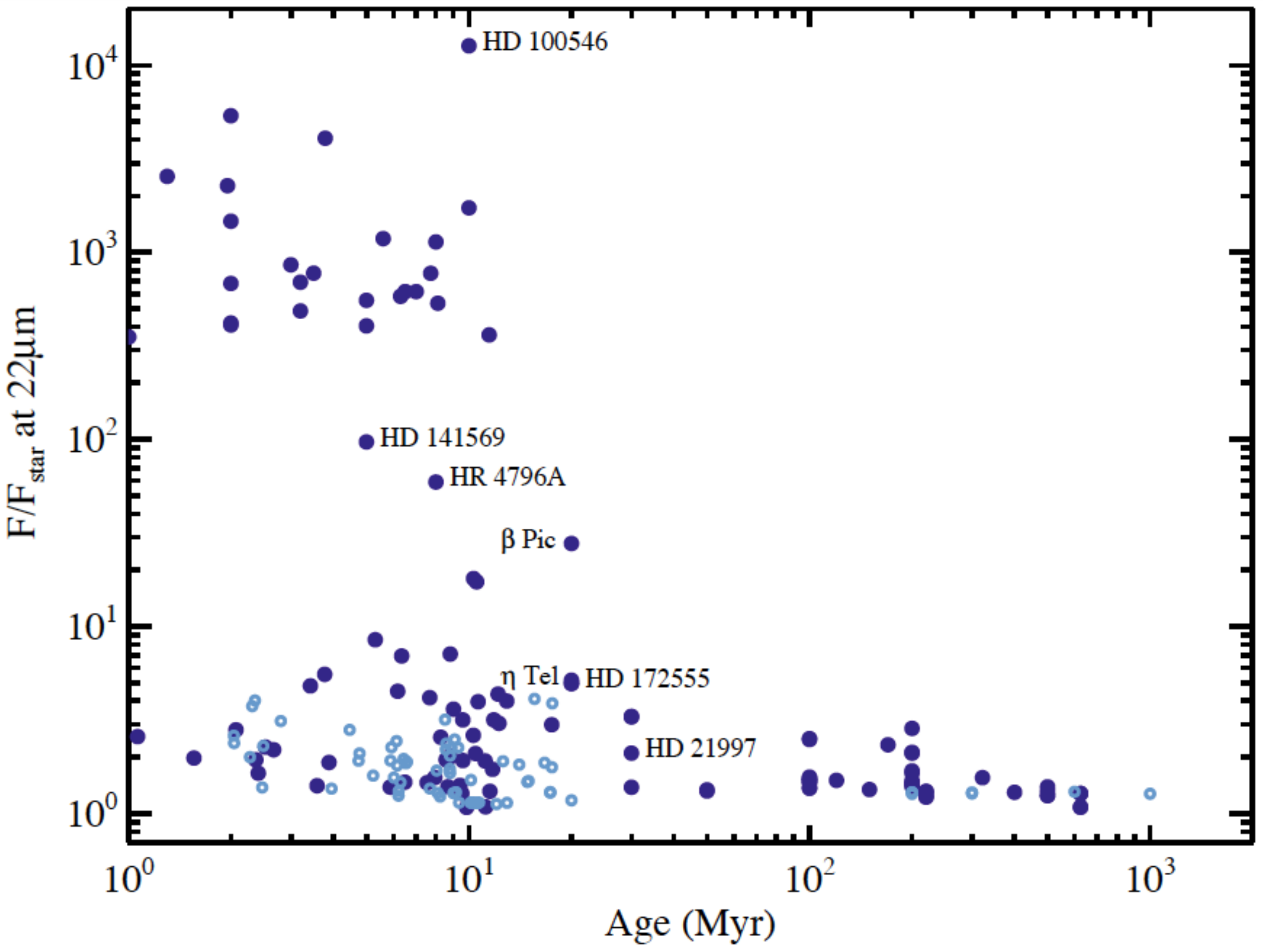} \\
\includegraphics[width=3.1in]{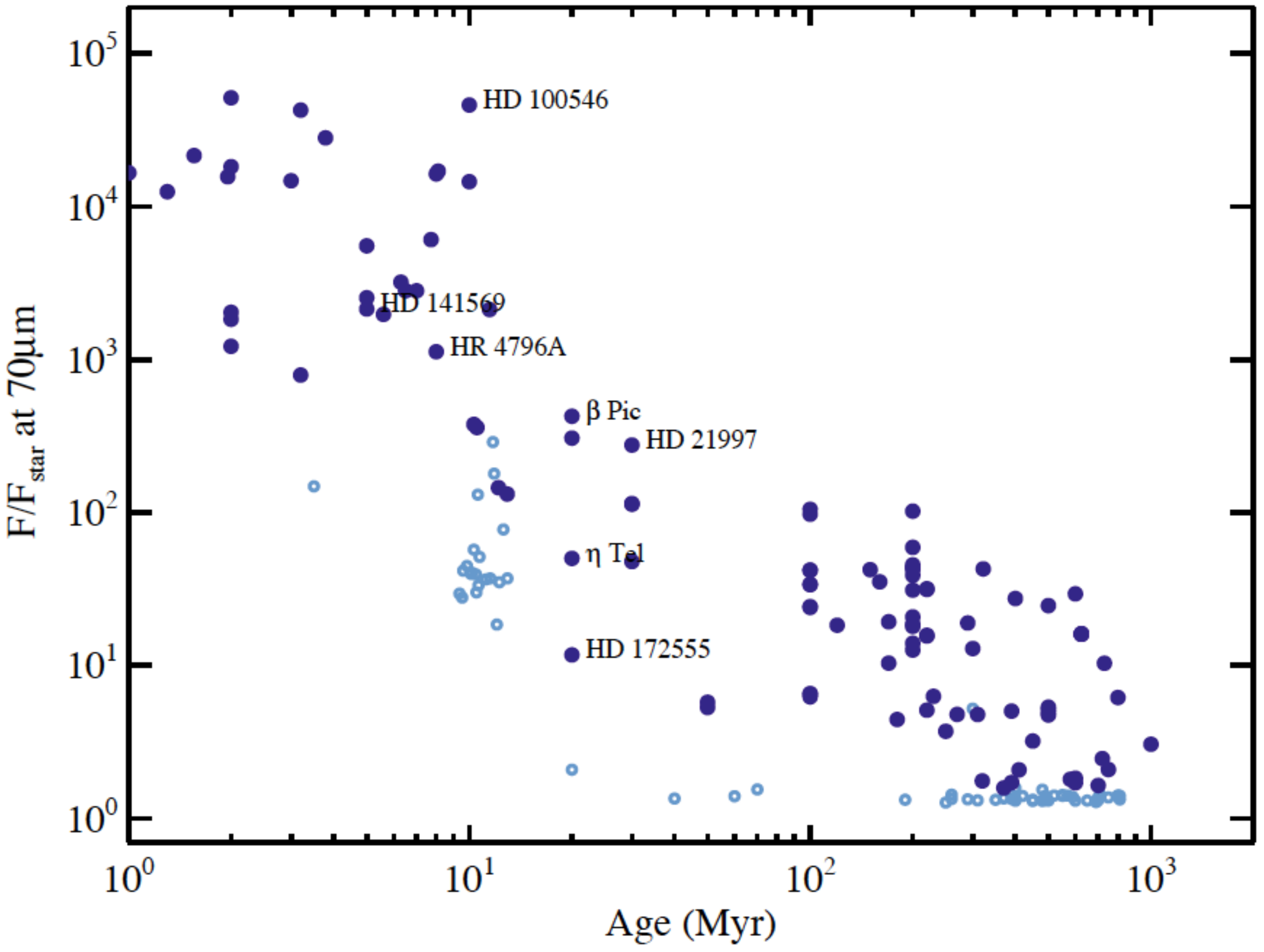} &
\includegraphics[width=3.1in]{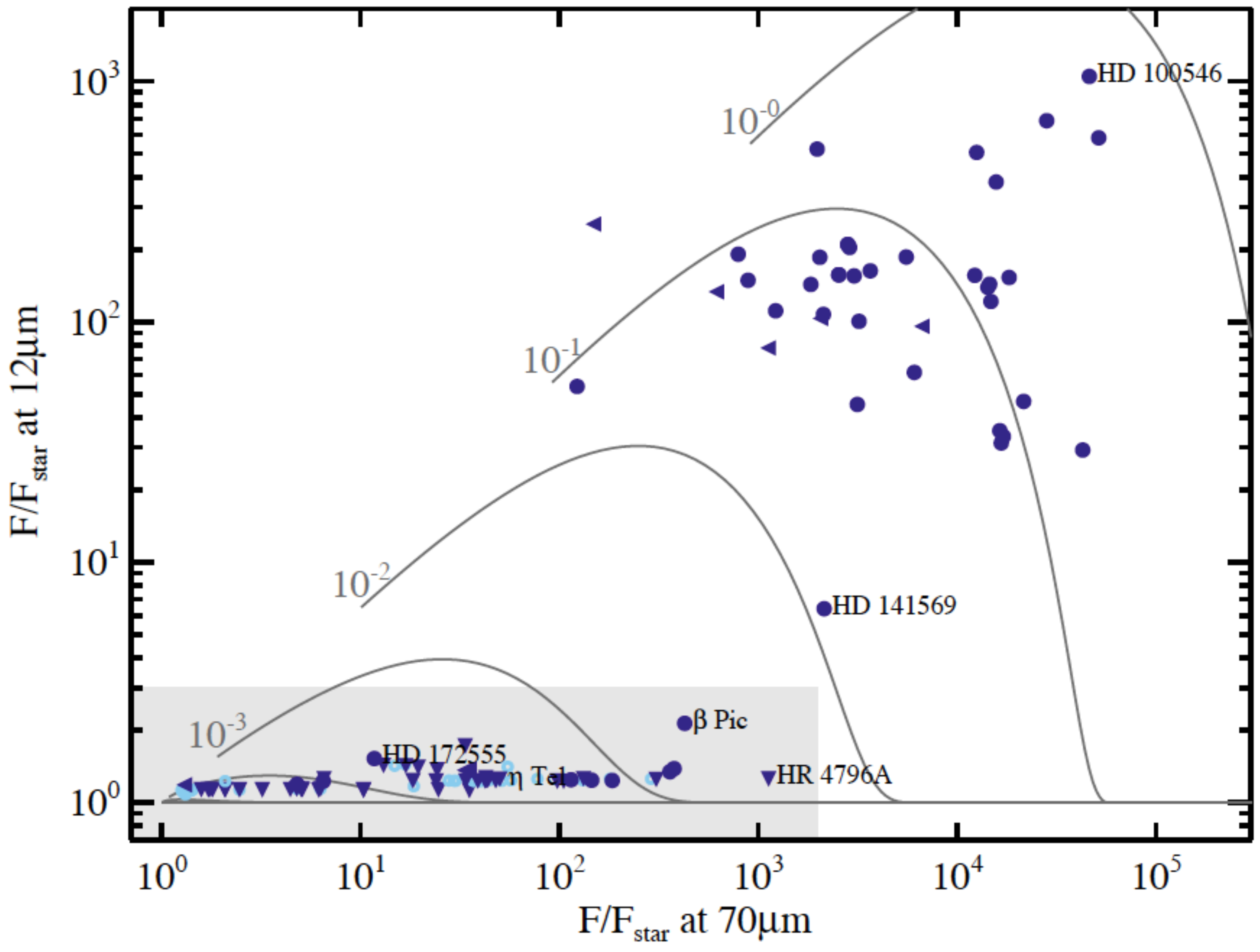} 
\end{tabular}
\caption{The ratio of observed to stellar
fluxes, i.e. the fractional excess $R_\lambda$ where 1 represents detection
of the photosphere, of A stars as a function of age:
\textbf{(Top left)} and \textbf{(Top right)} -- 12 and 22\,$\mu$m excesses from WISE,
\textbf{(Bottom left)} -- 70\,$\mu$m excesses from Spitzer and Herschel.
For all stars the photospheric level at these wavelengths was estimated
from a photosphere model fit to shorter wavelength data.
Filled blue circles are detections; open light blue circles are upper limits.
HR4796A, $\eta$ Tel and HD21997 are upper limits in the top left panel.
The \textbf{(Bottom right)} panel shows 12\,$\mu$m versus 70\,$\mu$m
excess on which the grey shaded region summarises the empirical classification
proposed for the boundary between protoplanetary and debris disk;
triangles are upper limits at just one wavelength (left-pointing for upper limits at
70\,$\mu$m and down-pointing for limits at 24\,$\mu$m).
The grey lines show the fractional luminosity implied if the spectrum is assumed
to be a single temperature black body that matches the excesses at these
two wavelengths.}
\label{fig:2470}
\end{figure*}

Since the majority of both debris and protoplanetary disks are identified
in mid-IR and far-IR surveys it would be helpful if a classification scheme
could be devised based on those surveys,
for example, based on the fractional excess $R_\lambda$ (the ratio of total flux
from a system to that from the star) at an
appropriate wavelength $\lambda$ in the range $5-500$~$\mu$m.
In Fig.~\ref{fig:2470} we show the fractional excess
at 12, 22 and 70\,$\mu$m, both for the individual objects discussed
in \S \ref{ss:ppd} and \S \ref{ss:dd},
and for A stars with a much wider range of ages.
At young ages this sample comprises relatively nearby HAe stars from
the literature and more distant A-stars in young
clusters from \citet{Hernandez2005}.
HAe stars have been the subject of numerous observational 
studies since their first identification as a group by \citet{Herbig1960}, and subsequent 
studies of \citet{The1994}, \citet{Sylvester1996}, \citet{vandenAncker1998}, 
\citet{SylvesterMannings2000}, \citet{Malfait1998} revised the membership list 
to eventually include not only the stars associated with a known star-forming region, 
but also stars in isolation (see \S \ref{ss:ppd}).
We compiled a sample of A-type stars from these works, and for 
which ages have been determined.
Older stars were taken from \citet{Su2006} and the
DEBRIS and SONS surveys \citep{Matthews2010,Panic2013}.
We also included sources with B9 spectral type designation, 
given that these are often characterised as A0 or B9/A0 stars.
We do not claim the plotted star sample to be free from bias, since it simply
comprises those in the literature, however it does plot both protoplanetary and
debris disks in a consistent manner to facilitate comparison of these populations.

Fig.~\ref{fig:2470} shows that at all wavelengths there is a clear downward trend of
excess with age.
In particular the absence of large excesses at ages $\gg 10$\,Myr is real,
since many stars were observed for which such excesses would have been readily
detectable. 
It is less clear whether the absence of low excesses at ages $<10$\,Myr is real, because
there is a bias in that young stars are less abundant and so are usually
found at greater distance, which means that deeper observations are required to
measure the levels of fractional excess that are more common around older stars.
Observations of nearby main sequence stars typically reach the
calibration limit below which it is not possible to discern an excess
photometrically (which is at $R_\lambda$ of around $1.10$), whereas those of the
more distant younger stars are often limited by the length of integration
or by the background confusion level, in particular at 70\,$\mu$m \citep[see also][]{Carpenter2009}.

There is an additional complication in that samples of young A stars are biased
toward having large excesses, because such stars are identified by their emission lines and
their interpretation confirmed by virtue of their excess infrared emission.
That is, it is not possible to tell from this plot whether large excesses are a majority
or a minority in this age range because it is a biased sample.
This bias can be addressed by observing all A stars in given star forming regions
to determine the distribution of excesses.
Such studies performed at near-IR wavelengths suggest that a significant fraction of
5-10\,Myr A stars should have already lost their protoplanetary disks \citep{Hernandez2005}.
We included stars from \citet{Hernandez2005} in Fig.~\ref{fig:2470}, which confirms
that a large fraction of these young stars have observational characteristics similar
to older debris disks, and suggests a bimodal distribution of excesses at young ages.
However, we also note that small number statistics and potential environment dependent
biases remain an issue, and of course the plot still includes the biased HAe
populations.

The clearest observational distinction between what we call a protoplanetary
disk and what we call a debris disk is at 12\,$\mu$m (Fig.~\ref{fig:2470}, upper left).
The dividing line at this wavelength is at $R_{12}=3$, which is slightly below
HD141569 which has been called both (but is a protoplanetary disk with this definition).
To avoid the possibility of identifying an old system in which a single asteroid
collided in the inner regions with a protoplanetary disk, we impose a further requirement
of $R_{70}<2000$ to identify a system as having a debris disk.
This classification scheme is summarised in the bottom right panel of Fig.~\ref{fig:2470},
where it is worth reminding the reader that this empirical classification scheme
is derived based on a sample of A stars, and so cannot necessarily be applied to
other spectral types.
The lines of constant fractional luminosity on that plot (under the assumption
that the spectrum resembles a single black body) show that this classification
results in a boundary at a fractional luminosity between $10^{-3}$ and $10^{-2}$.
One question that remains is whether this classification scheme would
misidentify transition disks with very clean inner gaps as debris disks.
While we are not aware of any such systems, these may be absent from the HAe class,
because of the fact that classification as a HAe
star initially relied on a near-infrared excess (among other things) as an important
indicator of their youth \citep{The1994}.
Indeed such systems do exist around around young lower mass stars
\citep[e.g., PZ99;][]{Mathews2012,Zhang2014}.

Note that in \S \ref{ss:phys} we will conclude that gas density is the most 
important physical distinction between the two classes of object, but the difficulty
of detecting this component, and the large uncertainties involved in deriving a
mass from the observations \citep{Miotello2014},
mean that we are essentially using the dust thermal emission
as a diagnostic of the presence of gas.
In this respect it should be pointed out that any classification scheme that relies only
on the dust emission spectrum must be fallible if there is a regime for which
the SEDs of protoplanetary and debris
disks are indistinguishable.

\subsection{Physical distinction}
\label{ss:phys}
Any observational classification for the debris disk -- protoplanetary
disk boundary should also consider the physical difference between
the two types of object.
The physical distinction that is usually invoked in the literature is that
the dust in debris disks is short-lived and so must be secondary in nature
\citep[i.e., continually replenished in the break-up of larger objects,][]{Backman1993}.
A short dust lifetime can be inferred if the dust is determined to be small
(microns in size) either from the thermal emission (e.g., from its temperature
if its radial location is known, or from mid-IR spectral features or the
sub-mm slope), or from the inferred scattering properties of the dust.
Poynting-Robertson drag removes such small dust on Myr timescales that are much shorter
than the age of the star precluding the possibility that the dust is primordial.

Such small, and likely secondary, dust also exists in protoplanetary disks, however the young age of
their host stars complicates the timescale arguments.
Radiation forces are also reduced by the high optical depth of protoplanetary disks,
meaning that only a negligible fraction of dust is directly exposed to stellar light
in the surface layer.
Rather the dust evolution is dominated by growth in collisions that is concurrent with
(and aided by) the dust sinking toward the midplane due to gravity,
and so to deeper and denser layers where the starlight is not able to penetrate.

While the above reasoning suggests that small dust in protoplanetary disks can be primordial, in fact this is 
unlikely for the bulk of the dust inside the optically thick region of the disk, and especially the disk 
midplane.
There the high dust densities result in frequent collisions, and in these collisions the dust is
expected to grow into larger pebbles \citep{Takeuchi2005, Testi2014}.
Indeed observational evidence for grain growth has been found in a number of sources, starting
with mid-IR observations \citep{Bouwman2001, vanBoekel2003}, and then longer wavelength observations
showing that millimetre dust is abundant in the outer disks
\citep{Testi2003, Natta2004, Wilner2005, AndrewsWilliams2007a}.
Since collisions are expected to remove the micron-sized dust on short timescales (through grain growth),
and interactions with gas would also remove the pebbles on short timescales,
the continued presence of both components throughout the protoplanetary disk phase
implies that collisions result in dust destruction as well as growth \citep{Dullemond2005},
and this is expected given the range of dust sizes and relative velocities anticipated in
the disk \citep{Brauer2007}.
Therefore, there is strong evidence that the bulk of the small dust we see in protoplanetary
disks is secondary in nature (i.e., it must be continually replenished from the destruction of larger grains),
just as it is in debris disks.

If the size distribution in a protoplanetary disk is a quasi-steady state in
which growth and destruction at each size are balanced
\citep{Windmark2012,Garaud2013}, then the main
physical difference with a debris disk could be that in the latter the collisional
velocities are much higher so that dust growth is no longer possible.
Furthermore it is tempting to identify the presence of gas with that distinction, because
gas damps collision velocities of the small dust (and big planetesimals) in
protoplanetary disks.
However, recent models have considered the possibility that a debris disk
started with a low level of stirring could persist for Gyr at detectable levels
even in the absence of gas \citep{Heng2010,Krivov2013}.
Indeed the observable properties predicted for unstirred disks (i.e., a spectrum
and resolved radius compatible with black body grains) match those of the cold
debris disks discovered by Herschel \citep{Eiroa2013,Krijt2014},
though the existence of this latter population is debated \citep{Gaspar2014}.
The sharp outer edge of the HR4796A debris ring in scattered light has also
been attributed to a low level of stirring \citep[i.e., low collision velocities;][]{Thebault2008}.

Such unstirred debris disks share many features in common with protoplanetary disks,
since neither require the presence of objects larger than the cm-sized pebbles
required to explain the mm-wavelength observations, though the unstirred debris
disks go further to imply an absence of planets (which would otherwise stir them,
unless the planets are on perfectly circular orbits)
which is not the case for protoplanetary disks.
In constrast, planetesimals are required in conventional debris disks to replenish
the small short-lived dust, the presence of which distinguishes them
from their unstirred counterparts, over main sequence lifetimes.
It is also likely that the stirring of planetesimals in debris disks is somehow caused by
the growth of planet-sized objects in their systems, either through the
gravitational perturbations of embedded Pluto-sized objects \citep{Kenyon2010}
or the secular perturbations of fully formed planets which need not
lie close to the disk in semimajor axis \citep{Mustill2009}.
Indeed a warp at 80\,au in the $\beta$ Pic debris disk provides
evidence that it is being stirred by a $9M_{\rm{Jup}}$
planet seen to be orbiting at 9\,au \citep{Augereau2001,Lagrange2010}, and 
spiral structure in the HD141569 disk \citep{Clampin2003} may have a similar
implication \citep{Wyatt2005}.
However, planets are not required in debris disks since planetesimals could simply
be born on orbits with high collision velocities \citep[e.g.,][]{Walmswell2013},
or achieve them as a natural outcome of disk evolution
\citep[e.g., due to asymmetric mass-loss;][]{Jackson2013}, as damping
timescales can be longer than gas dispersal timescales for large enough planetesimals.

Clearly the level of stirring that planetesimals receive and their radial distribution
depend on the architecture (and indeed existence) of the planetary system within which
they reside.
An important question then is when systems of (putative) planets and planetesimals achieve
the structures seen around main sequence stars.
The ages of CAIs (Calcium-aluminium-rich inclusions) and meteorites in the
Solar System imply that formation of
planetesimals (at least in the inner Solar System) occurred early on in the protosolar
nebula \citep{Shukolyukov2003} and over an extended period of $>1$\,Myr
\citep{Wadhwa2005,Cuzzi2008}, though whether those planetesimals were born with sizes of
100s of m or 100s of km remains debated \citep{Morbidelli2009,Kobayashi2014}.
It also seems likely that the formation of planetary embryos (if they form at all) precedes the
transition disk phase, since it makes more sense for these objects to form in the long period when 
sufficient mass was available, rather than during the short transition phase
during which that mass was depleted.
Moreover there is some observational evidence for planets or companions in transition disks
\citep{Huelamo2011, Kraus2012, Quanz2013, Biller2014, Reggiani2014}, and the
gas giant planets seen around many main sequence
stars must have formed in the presence of significant quantities of gas.

Hence a useful hypothesis to test is that the planetary system and
its planetesimal belts are already largely in place before the transition disk phase.
This does not rule out that the planets continue to grow at later phases through collisions,
or that they migrate in the gas-rich protoplanetary disk phase \citep{Walsh2011},
and neither does it preclude the possibility of a period of dynamical restructuring
following the dispersal of the protoplanetary disk \citep{Gomes2005,Raymond2012}.
Such system changes would simply weaken the extent to which this hypothesis is true.
If these changes commonly render systems unrecognisable between their protoplanetary and debris
phase architectures then the hypothesis is no longer useful, and this would be the case
if for example these structures were only formed during disk
dispersal \citep[see \S \ref{ss:mm}; e.g., ][]{Alexander2007}.
As long as this is not the case, we should be looking to debris disks to tell us
about the distribution of the less observable components of protoplanetary
disks (i.e., the planets and planetesimals that are embedded in a much brighter
disk of gas and dust).
Furthermore, this hypothesis allows us to propose a clear physical distinction
between protoplanetary and debris disks, which is that protoplanetary disks
contain a large quantity of primordial gas which is also suffused with
small dust grains, made up of the dust which did not yet grow to planetesimal
sizes, as well as dust released in planetesimal collisions which is then radially mixed
through entrainment in the gas.
It then makes sense to peg the level of gas at which the transition is defined as that which
is sufficient to dominate the motion of the dust, since that may be empirically
determined from its effect on the observational properties of the dust disk, and
so provides some justification for basing the observational
distinction given in \S \ref{ss:obs} on observations of the dust rather than
the gas.

\section{Five stages in the transition}
\label{s:stages}
Given that we witness individual disks at just one evolutionary stage and
only have incomplete information about each of them, and moreover stellar age
is not a reliable measure of the degree to which an individual disk has evolved,
it is always going to be difficult to make firm conclusions about the evolution through the
transition from the protoplanetary to debris stage.
This is compounded by the rapidity of the transition which leaves us with 
few objects that are truly caught in transition.
The following discussion is thus somewhat speculative, but is based on what
can be learnt from looking at disks as close to the transition
as possible and noting the quantity of different components that are present.
We have identified five key stages in the evolution, represented by the
young disks discussed in \S \ref{ss:ppd} and \S \ref{ss:dd}.
However, we are not claiming that any of these
disks will evolve into systems that resemble each other, and neither
do we claim that the order of the evolution is uniquely determined;
indeed these stages may take place concurrently rather than
consecutively, similar to the homologous depletion proposed for protoplanetary disks
\citep{Currie2009}. 
Rather we expect the evolution to be inherently chaotic, presumably
determined by how planet formation and disk dispersal processes play out in each system
(see \S \ref{ss:phys}).
Nevertheless, the subdivision of the evolutionary stages in this way provides
a framework for discussion from which to make progress in our understanding
of the transition.

\begin{figure}[tb]
\includegraphics[width=3.2in]{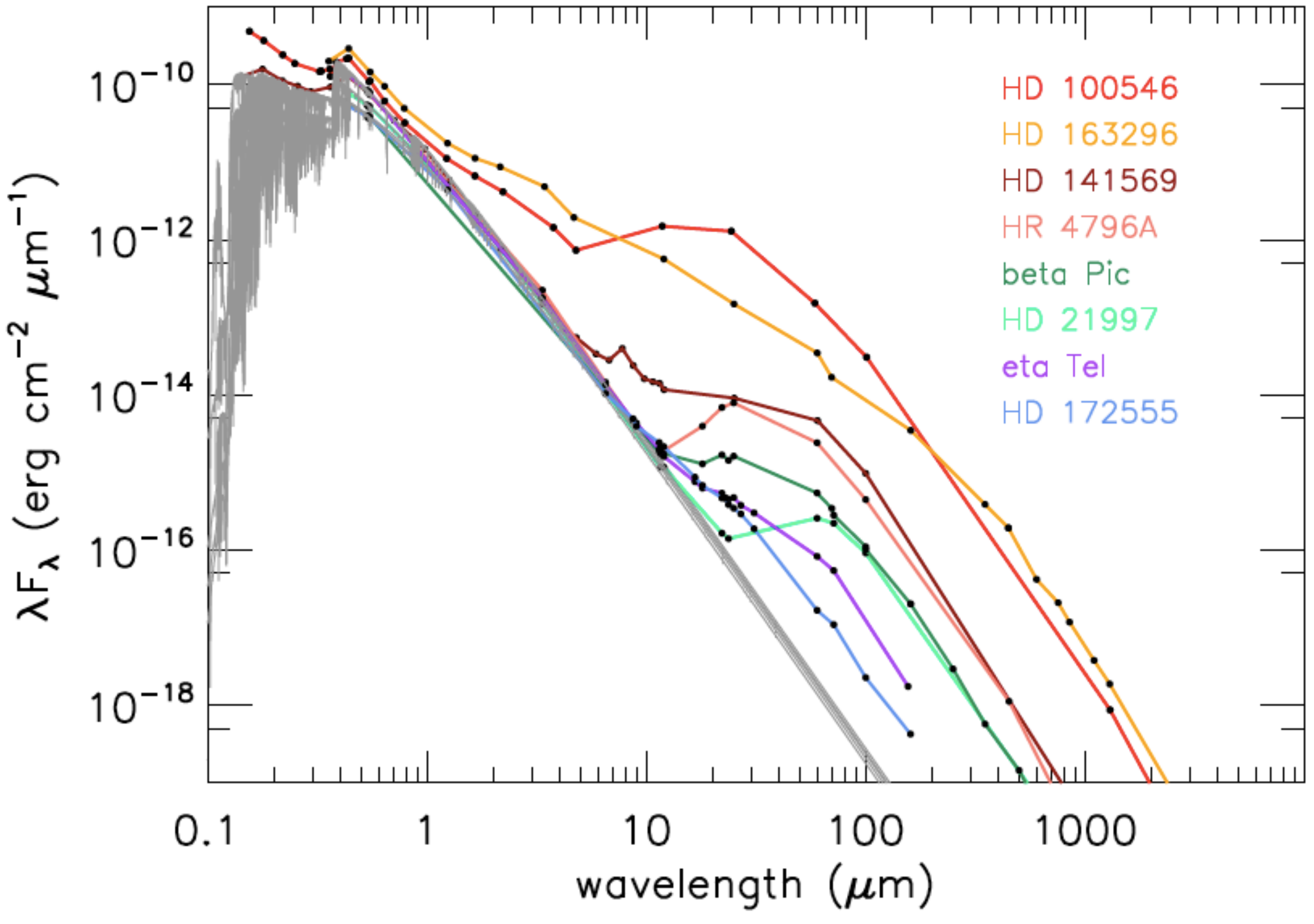}
\caption{Spectral energy distributions of A star disks selected for
the similar luminosities and effective temperatures of their host stars.
All emission spectra have been scaled to 100\,pc
so that the photospheric emission for all should be similar to
that of HR4796A which is shown in grey.
Broad band photometry for each star is shown with black dots without error
bars for clarity.
The spectra obtained by joining these photometric points illustrate the evolution
from transition disk (HD100546) to one depleted in mm-sized grains
(HD141569) to one in which the inner regions have been cleared and
the remaining planetesimals are left in a narrow ring (HR4796A).
Alternative outcomes (or the descendants of such rings) include
$\beta$ Pic and HD21997 which have only slightly lower cold dust levels,
while $\eta$ Tel and HD172555 represent
outcomes in which cold dust is more depleted but (relatively) small quantities
of dust exist in the inner regions which may be a transient phenomenon.
}
\label{fig:sedevol}
\end{figure}

We start in \S \ref{ss:transition} with the transition disk stage illustrated
by HD100546.
This is followed in \S \ref{ss:mm} by the depletion of mm-sized dust in the outer
disk illustrated by HD141569, and a discussion of the removal of the dust in the inner regions in
\S \ref{ss:hot} the continued existence of which is illustrated by HD172555 and $\eta$ Tel.
The removal of the gas is discussed in \S \ref{ss:gas}
using the examples of $\beta$ Pic and HD21997, 
and finally the ring-like concentration of planetesimals illustrated by HR4796A
is discussed in \S \ref{ss:ring}.
The SEDs of a few of these objects are shown in Fig.~\ref{fig:sedevol}, which
illustrates this transition, and is similar to figure 8 of \citet{Kennedy2014}
which shows the evolution of the circumstellar disks of Sun-like stars.

\subsection{Transition disk}
\label{ss:transition}
The presence of inner holes in transition disks has a profound effect on
their near infrared spectrum and their inner disk structure revealed in high resolution imaging, 
but otherwise transition disks are not significantly different to full disks
in terms of their far infrared luminosity \citep{Rodgers-Lee2014}.
While this class may correspond to the low end of the protoplanetary disk
dust mass distribution \citep{OwenClarke2012}, a significant number of massive
transition disks also exist (such as HD100546).
However, transition disks may have larger dust grains than
full disks \citep{Pinilla2014}, as well as lower levels of OI gas which could indicate that they are
less flared \citep{Keane2014}, suggesting they may be more evolved.

The fraction of HAe disks that are classed as transition disks is poorly known
for the reasons described in \S \ref{ss:ppd}.
However, the fraction of the protoplanetary disks found around lower mass stars that are classed
as transition disks ranges from 10-20\% \citep[e.g.,][]{Muzerolle2010, Furlan2011}.
If the same fraction applies to HAe disks one implication is that,
since the remaining 80-90\% of full disks in these regions may also pass through a transition disk phase,
the fact that only a few percent of stars are found to have giant planets
at large distances \citep[e.g.,][]{Biller2013}
suggests that planet-induced clearing is unlikely to explain all 
transition disks unless the planets required to do the clearing can be below the detection
threshold of the imaging surveys (a few $M_{\rm{Jup}}$).

On the other hand planets (or indeed multiple planets) provide a mechanism to 
carve a hole while maintaining accretion at levels observed toward some transition disks
\citep{Williams2011, DodsonRobinson2011, Zhu2012},
and a planet does not necessarily have to be massive to open a gap \citep[][]{Baruteau2013}.
Furthermore, in some transition disks a lop-sided asymmetry in millimetre dust has been found
\citep{vanderMarel2013,Casassus2013}, providing indirect evidence for trapping of millimetre dust in 
planet-induced pressure waves \citep{Pinilla2013}.
The alternative mechanism invoked to explain the inner clearing is photoevaporation,
which is indeed expected to result in disk holes \citep[e.g.,][]{Alexander2007},
although this is a less viable explanation for the most massive transition
disks, since this mechanism operates once disk mass is already depleted \citep{OwenClarke2012},
and also has problems explaining the accreting transition disks \citep[e.g.,][]{Cieza2010}.
A realistic picture probably includes both processes;
e.g., clearing by planets very close to the star could trigger efficient photoevaporation
such that transition disks do not necessarily have planets close to the outer edge of the hole or gap.

Considering that the inner few tens of au of most debris disks are also devoid of material, 
it is tempting to look at transitional disks as one of the steps towards debris disk formation. 
However, the strong coupling of the dust in transition disks to the distribution of gas
(which is no longer present at later phases) means that there is no guarantee that the dust
in transition disks traces the distribution of planetesimals.
Indeed, it is possible that any signatures of the hole that existed in the transition disk and its
location may be erased once the gas disperses.

\subsection{Depletion of millimetre-sized dust in the outer disk}
\label{ss:mm}
As illustrated in Fig.~\ref{fig:sedevol} the most easily detected observational
difference between protoplanetary and debris disks is the 
excess emission at a range of wavelengths, from infrared to the millimetre.
In this section, we focus on the evolution of the largest observable dust particles traced
by millimetre-wavelength observations.
Millimetre emission arises predominantly from dust a few hundred microns to a few centimetres
in size, and is often used as a dust mass tracer because it is optically
thin, contrary to the case of shorter wavelengths.\footnote{Here optical depth refers to that
along the line-of-sight to the observer at the wavelength in question.
In the optically thin regime the observer sees all of the emission from the disk at this
wavelength, whereas much of that emission remains hidden in the optically thick regime.}
Evidence of grain growth to millimetre sizes exists even at the early
stages of protoplanetary disk formation, and arguably this is the size range at which the majority of the solid
mass in a protoplanetary disk resides throughout its evolution (though a large mass in smaller and larger
grains can be hidden by optical depth and opacity effects, respectively).
The conclusions reached about the evolution of the population of millimetre grains may
broadly apply to that of the population of smaller grains, since these are linked to some extent
through the size distribution (see \S \ref{ss:phys}), with the caveat that the different forces
acting on dust of different sizes through its interaction with gas and stellar radiation adds
complexity to such inferences that will not be discussed here.

\begin{figure}[tb]
\includegraphics[width=3.2in]{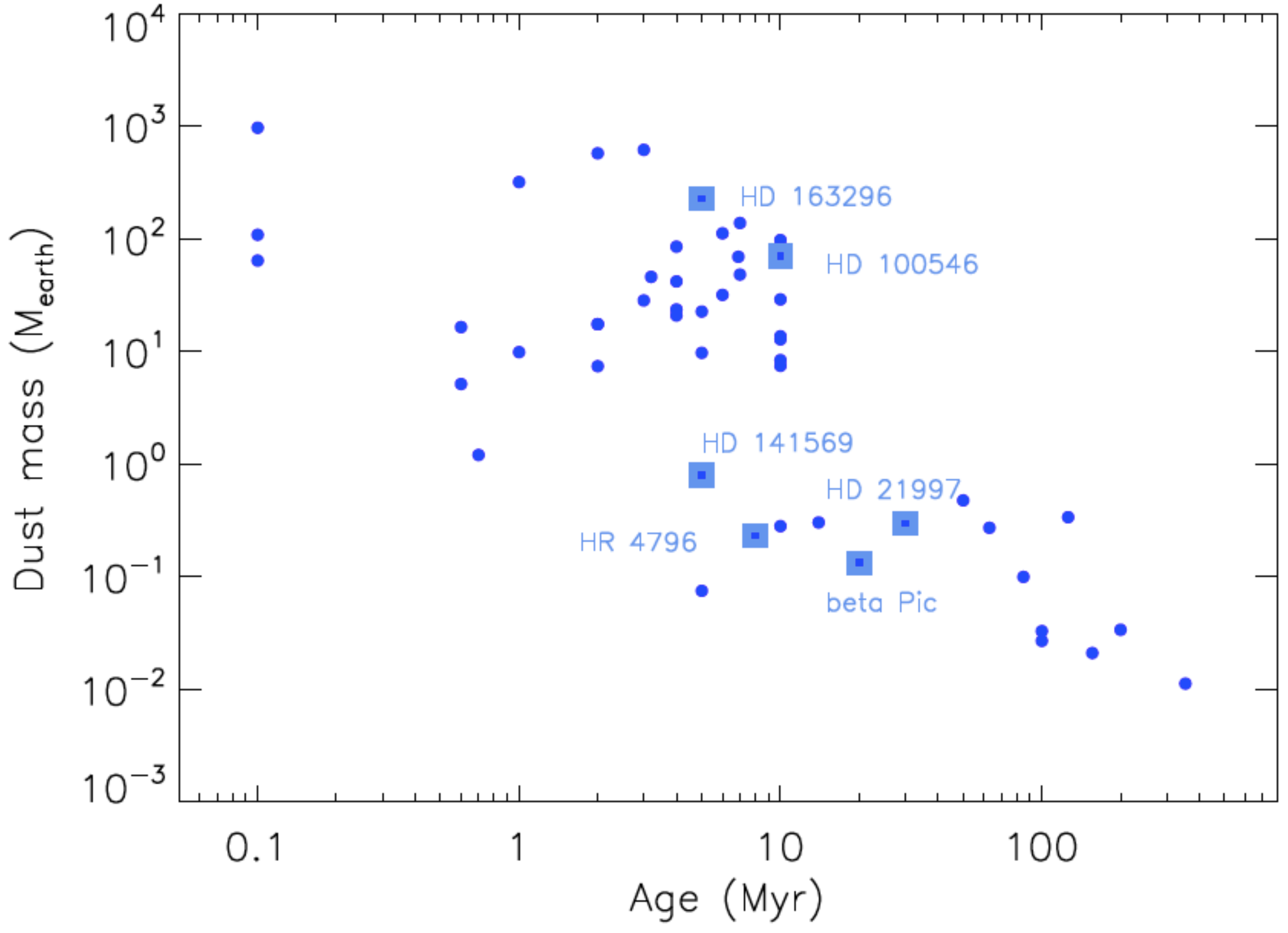}
\caption{Mass in millimetre-sized dust grains inferred from sub-mm observations
as a function of stellar age for A and B-type stars, adapted from \citet{Panic2013}, including also
masses from \citet{Sandell2011} and \citet{Thi2013a}.
Uncertainties in dust opacity dominate the uncertainties in the mass estimates,
which could be up to 10 times higher or 2 times lower. 
Absolute ages are only known to within a factor of a few, though relative ages
within the sample should be more accurate.
}
\label{fig:submm}
\end{figure}

The evolution of dust mass around nearby stars is shown in Fig.~\ref{fig:submm} \citep{Panic2013}.
While the interpretation of this plot is complicated without a detailed treatment
of the non-detections \citep[for a consideration of this for protoplanetary
disks around lower mass stars see][]{Andrews2007}, it is clear that there is a neat division at
$\sim 10$\,Myr between younger (protoplanetary) disks for which there is $>1M_\oplus$ of dust,
and older (debris) disks having $<1M_\oplus$ of dust;
old stars with $>1M_\oplus$ of dust would have been readily detectable so their absence
is independent of considerations of the non-detections that are not included on this
figure.
The fact that the transition disk of HD100546 lies firmly in the protoplanetary disk mass
range is part of the reason that \S \ref{ss:transition} stated that
such systems have not progressed very far along the evolutionary track to losing
their protoplanetary disks.

While it is not possible to generalise into a sequence from individual objects,
and we should reiterate the caveat that the order and consecutive nature
of the stages is uncertain,
the loss of the mm-sized grains as the next stage in the process
is guided by HD141569, a 5\,Myr HAe star which has all the characteristics of a protoplanetary
disk (10$^{-4}$~M$_\odot$ of gas, PAH emission, spatially 
extended scattered light features), but has a dust mass of $0.7M_\oplus$ that
is significantly depleted compared with other protoplanetary disks (see \S \ref{ss:ppd}).
The depletion of mm-sized grains likely occurs quickly given the lack of evidence for
evolution in their mass during the protoplanetary disk phase \citep[e.g.,][]{Carpenter2014},
though deeper mm-wavelength observations would help to constrain the evolution
in this phase.

The fate of millimetre-sized dust grains in protoplanetary disks is uncertain.
One possibility is that the concentration of such material into radially narrow
features following the carving of an inner hole in the transition disk
\citep{vanderMarel2013} results in high
enough densities for gravitational collapse or streaming instabilities and the
direct formation of planetesimals \citep{Chiang2010, Johansen2014}.
A similar fate is suggested by models in which the gas in protoplanetary disks is
dispersed from the inside out from photoevaporation by stellar UV and X-ray radiation 
\citep{Clarke2001}, since dust in such models gets pushed out
with the photoevaporation front which sweeps up the dust
\citep{Alexander2007}.
However, neither of these models can follow the evolution of the dust up to the formation
of planetesimals, because their assumptions break down once dust densities
exceed gas densities.
Thus it is also possible that these grains decouple from the gas before they have
had a chance to grow into planetesimals and perhaps are
deposited as a ring of debris at the location where this occurs.
Such debris rings could then be depleted by rapid collisional erosion if they have also been excited
to high enough collision velocity by this stage.

A more general consideration of this problem shows that
there are just two outcomes for the mm-sized grains (recalling that these likely comprise
a significant fraction of the solid mass of a protoplanetary disk).
Since they cannot be dragged away by a photoevaporative wind, 
and P-R drag takes many 10s of Myr to deplete mm-sized dust from 10s of au,
their loss can only be through collisions.\footnote{The mm-sized dust can also be accreted
onto the star through drag induced by the gas, but since this process would also occur
throughout the earlier protoplanetary disk phase, relative to which the transition is
rapid, this likely only affects a small fraction of the dust.}
If those collisions are predominantly destructive
then their mass will end up in much smaller dust grains
that can be removed by various mechanisms, such as radiation pressure blow-out
or coupling to the photoevaporative flow.
The alternative is for the mass originally in mm-sized dust to end up (through collisions)
in large objects,
either through their growth into planetesimals, or through their accretion onto
already formed planetesimals or planets.
The continued presence of gas after the depletion of the mm-sized dust in HD141569
seems to argue against loss by collisional erosion, since that gas would damp collision
velocities and favour grain growth.
However, since we do not know yet exactly where the mm-sized dust lies in relation to
the gas in HD141569 \citep[e.g., other transition disks with low millimetre fluxes
have been found to arise from systems with small outer disk radii;][]{Pietu2014},
it may be that these two components are sufficiently disconnected for
collisional erosion to be viable, a possibility that will be probed with millimetre
interferometric observations (Flaherty et al., in prep.).

The growth of the mm-sized dust into planetesimals on the other hand makes
sense if one was to believe that solid mass is conserved between the protoplanetary and debris disk
phases (as is the assumption for example in the minimum mass solar nebula model).
While the dust mass has undeniably been depleted by the debris disk phase (see Fig.~\ref{fig:submm}),
this must be the tip of the iceberg, since if all of the observed solid mass was all
there was, then it would be collisionally depleted on Myr timescales.
The presence of debris disks at 100s of Myr ages thus implies the existence of
planetesimals of at least km in size that are feeding the observed dust
\citep{Wyatt2002}.
This leads to inferred masses of 10s of $M_\oplus$ of solid material in the debris disks
that are comparable with those present in protoplanetary disks.

We do not favour either scenario here, but point out that the fate of mm-sized dust has significant
implications for our interpretation of young debris disks.
For example, if the mm-sized dust (which is seen to persist into the transition disk phase,
\S \ref{ss:transition}) grows into planetesimals then the debris disk architecture is
not already in place during the protoplanetary disk phase, invalidating the hypothesis proposed
in \S \ref{ss:phys}.
Conversely if the mm-sized dust grains are collisionally depleted then the young bright debris
disks (like HR4796A) could be an ephemeral phenomenon associated with the erosion
of this population, which 
is superimposed on a longer term (fainter) evolution of any pre-existing planetesimal belts.
Either way, this emphasises that there are two potential sources for dust in debris disks:
pre-existing planetesimals, and the destruction of whatever structure the mass in
mm-sized dust got shaped into during disk dispersal.
It will be hard to discern these sources observationally.

\subsection{Evolution of hot dust in inner regions}
\label{ss:hot}
Hot dust in the inner regions of protoplanetary disks is a natural consequence 
of models in which the dust is well mixed with the gas.
Even if the dust which started in the inner few au has long since grown into planetesimals or
planets, or been accreted onto the star, this region is continually repopulated by dust
which started off further out but then migrated in either through gas drag or simply 
with the bulk motion of the gas.
Here we point out that another source for hot dust in protoplanetary disks
is in debris-like processes (i.e., in processes more commonly associated with debris disks,
as opposed to the dust growth and drift processes considered to dominate in a protoplanetary
disk).
This may particularly be the case in systems like HD100546
where we know the hole is devoid of gas (both H$_2$ and CO),
but contains hot dust at $<0.7$\,au.
Such \textit{pre-transition disk} systems are often considered to arise at the onset of
photoevaporation which carves an annular gap at 10s of au, whereupon the inner
disk drains onto the star \citep{Clarke2001}.
However, it is worth considering whether this dust is second generation, formed from
collisions of larger bodies which may also have been present earlier on.
Here we focus on the explanations that have been proposed for the origin of hot dust
seen in debris disks, noting that it may be productive to consider whether these mechanisms
are also operating at earlier phases \citep[e.g.,][]{Kennedy2014}.

Three of the six 8-20\,Myr A stars have hot dust.
In the case of HD172555 it only has hot dust, at a temperature that puts
it at a few au, a location confirmed with mid-IR imaging and interferometry
\citep{Smith2012a}.
For $\eta$ Tel there are two dust components, one at 24\,au that is resolved
in mid-IR imaging \citep{Smith2009}, and another at a few au that is unresolved
in those images and also inferred from the spectrum.
The question for these hot components is: are these asteroid belt analogues,
sublimation of comets scattered in from outer regions, or evidence for the ongoing
formation of terrestrial planets or super-Earths?

A growing number of two temperature disks found around older (100s of Myr) stars
\citep{Morales2011,Su2013,Chen2014},
i.e., systems with both cold and hot dust like $\eta$ Tel,
gives an indication that the hot dust in some of these
systems may have a cometary origin \citep{Kennedy2014b},
since the outer belts provide an appropriately long-lived source population.
A rather contrived set of planetary system properties would be needed
to result in detectable hot dust from an outer planetesimal belt that lies
below the detection threshold \citep{Bonsor2012},
and so this is unlikely to be the origin of hot dust-only systems like HD172555.
However, this is a reasonable proposition for young systems with hot and cold
dust, particularly if the early phase in the system's evolution is characterised
by dynamical settling and so intense cometary activity \citep{Bonsor2013a}.
Indeed it has been proposed that transient spectral absorption events in some transition
disks including HD100546 is caused by Falling Evaporating Body (FEB)-like
cometary activity \citep{beust2001}.

High levels of hot dust is almost exclusively a phenomenon of young ($\ll 100$\,Myr)
stars \citep[see Fig.~\ref{fig:2470};][]{Melis2013,Kennedy2013}.
This is consistent with both an asteroid belt and a planet formation interpretation,
since rapid collisional depletion at such proximity to the star means
that asteroid belts quickly evolve below the detection threshold \citep{Wyatt2007a}
and rapid collisional evolution is also the explanation for the eventual cessation of
planet formation processes.
The formation of terrestrial planets or super-Earths is an attractive proposition for
the origin of the hot dust because models of such processes that are successful at
reproducing the Solar System's terrestrial planets predict that hot dust should
be readily detectable at this epoch.
Such models start out with km-sized planetesimals which grow through 
collisions into embryos, and later the number of embryos is whittled 
down as they coalesce into a small number of roughly Earth-mass planets.
Planetesimal collisions would be expected to produce dust, the level of which
is predicted to be at an easily detectable level
up to several 10s of Myr \citep{Kenyon2005}.
Since such models do not include the debris created in giant impacts between
the embryos, the aftermath of which should also be detectable for $\sim 15$\,Myr
\citep{Jackson2012}, it is clear that if planet formation in this manner
(whether the outcome is 1 or 10 Earth mass planets) is ongoing
after or indeed during the protoplanetary disk phase, then it should be detectable.
There may be evidence from the mid-IR spectrum to support this for
HD172555, since a sharp feature at 11~$\mu$m is attributed to silica
that is expected to be produced in high velocity collisions
such as those between planetary embryos \citep{Lisse2009,Johnson2012}.

Given the plethora of opportunities for creating dust in the inner regions of
these systems it is not a surprise that it is detected around many young
main sequence A stars.
It would only be absent if the inner regions were completely empty of
planetesimals, any rocky planets present were already sufficiently separated for further
collisions amongst them to be unlikely, and the outer planetary system was arranged
so as to prevent planetesimals from any outer planetesimal belt being scattered
into the inner regions.
Moreover, unless the formation of asteroid belts or terrestrial planets occurs during
the transition, these objects must be
present during the protoplanetary disk phase (including the transition disk phase).
As such it is important to question
how much of the hot dust in protoplanetary disks of all varieties arises from such
processes.
However, the empirical observation in \S \ref{ss:obs} that large levels of 12\,$\mu$m excess
(which were quantified as $R_{12}>3$) are only present in protoplanetary disks could indicate that debris-like processes
can only supply a limited level of excess, and that gas must be present for higher
levels (e.g., by aiding the radial transport of dust or by slowing down dust removal
processes), which therefore provides some physical motivation (\S \ref{ss:phys})
for the observational classification proposed in \S \ref{ss:obs} for the
protoplanetary-debris transition.

\subsection{Disappearance of gas}
\label{ss:gas}
\begin{figure}[tb]
\includegraphics[width=\columnwidth]{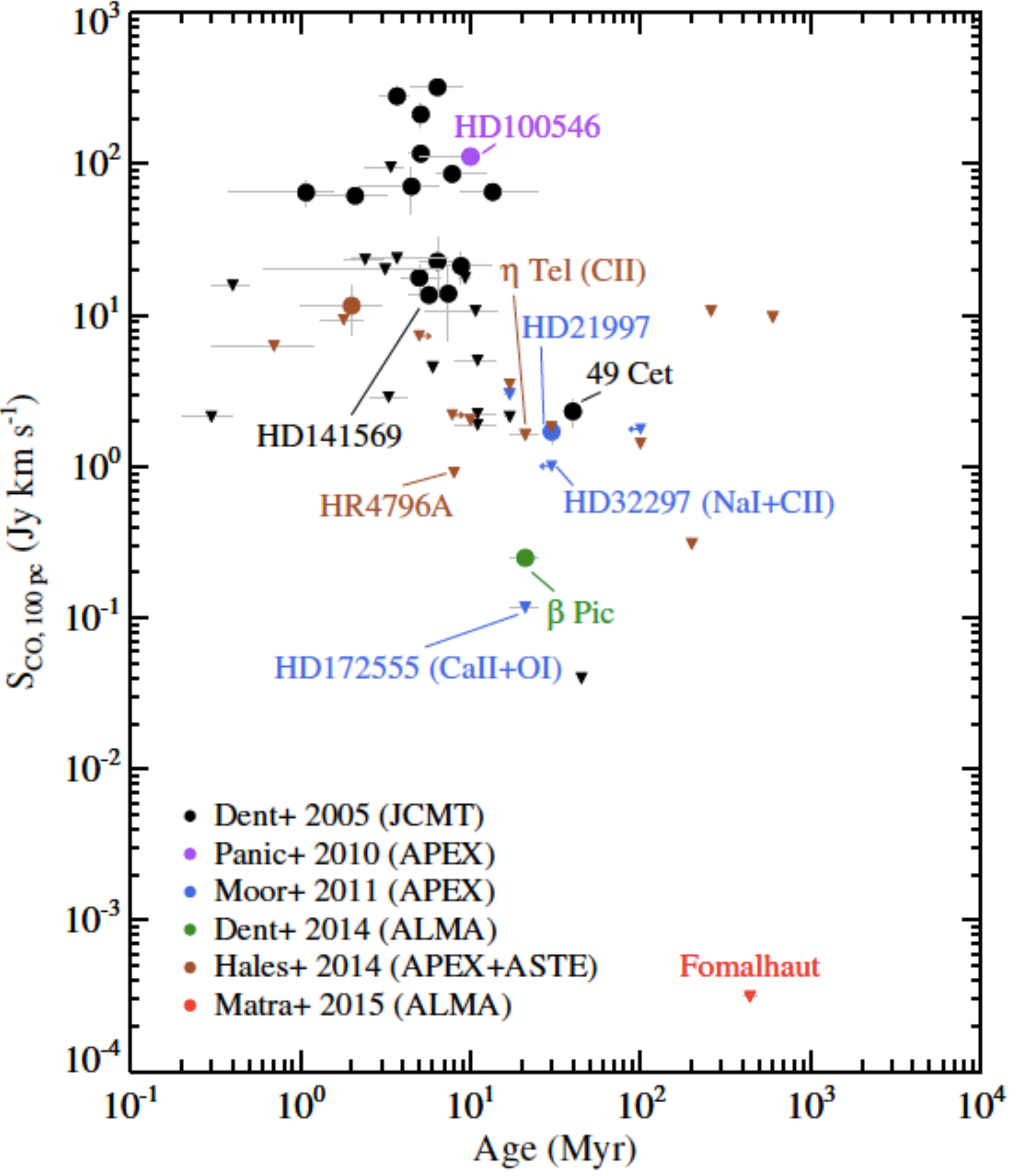}
\caption{Integrated intensity in the CO $J=3$--2 line scaled to a distance of 100\,pc
as a function of stellar age for A and B-type stars with observations reported in the literature
\citep[i.e., including more stars than the samples in \S \ref{s:diff}; adapted from][]{Matra2014}.
Filled circles are detections (coloured by their reference of origin as noted in the
figure), and triangles are upper limits.
Uncertainties are noted with grey lines, and are less than a factor of two in
integrated intensities, while ages have similar uncertainties to those reported for
Fig.~\ref{fig:submm}.}
\label{fig:gas}
\end{figure}

Measuring gas masses is extremely challenging in both protoplanetary and debris disks for 
different reasons.
In protoplanetary disks, the bulk of the gas composition is inherited from the 
interstellar medium and hence its main constituent is H$_2$.
However, even when H$_2$ is observed its emission only traces a minor fraction of the
total disk mass \citep{Thi2001, Carmona2011}.
While the second most abundant molecule (CO) is readily observed in HAe disks from the infrared
to millimetre, CO millimetre lines are a poor indicator of disk gas mass because they are
optically thick, though they are a useful diagnostic of the size and
geometry of HAe disks \citep{Panic2009}.
The situation is improved for the optically thin low-abundance isotopes such as C$^{18}$O,
however large uncertainties linked to the conversion to the total H$_2$ mass remain.
This is particularly true in the outer disk region where the bulk of the mass resides and where
low temperatures render CO invisible as it freezes out onto dust grains, but also in the
surface layers of the disk, as these may be affected by selective photodissociation
\citep{Miotello2014}.

In the case of debris disks, any gas that is detected may be secondary, 
i.e., released from large dust and planetesimals where it resides in the form of ice
\citep[e.g., ][]{Zuckerman2012}.
If so then the chemical composition of such gas would be markedly different from primordial gas,
likely dominated by water and CO followed by the more complex species often observed in comets
in the Solar system \citep[e.g.,][]{Mumma2011}.
Similar to protoplanetary disks the most comprehensive searches for gas toward debris disks 
involve searches for CO millimetre lines, and in Fig.~\ref{fig:gas} we show the evolution of
the integrated CO $J=$3--2 line intensity (scaled to 100\,pc) as a function of stellar age
including both classes of disk for all A and B-type stars for which observations are reported
in the literature \citep{Dent2005, Panic2010, Moor2011, Dent2014, Hales2014, Matra2014}.
We already acknowledged that this observable does not have a one-to-one correspondence with the 
mass of CO in protoplanetary disks due to optical depth effects, and this is also true for debris
disks but for different reasons.
In debris disks the low gas densities result in poor coupling
between the gas and the dust.
This means that the gas excitation mechanisms and the processes that drive its dissipation are
expected to be different from those in the higher density protoplanetary disk regime \citep{Matra2014}.
Nevertheless, even if the sample in Fig.~\ref{fig:gas} is significantly biased due to
the limited number of stars that have been observed,
it illustrates at what age and down to what level stars have been searched for this CO
transition, which can be viewed as a useful (if flawed) proxy for gas
mass, or at least as an indicator of significant gas levels.

Fig.~\ref{fig:gas} shows that it is generally the case that CO has not been detected around stars older than
10\,Myr or those classed as debris disks.
However, as acknowledged in \S \ref{ss:mm}, it is still present in some systems
which have a depleted level of mass in mm-sized dust like HD141569 \citep{Zuckerman1995,Dent2005}.
Moreover, there is a growing number of young debris disk systems for which CO has
been detected.
The most recent of these is $\beta$ Pic for which CO had previously been
detected along the line-of-sight to the star \citep{Roberge2000,Troutman2011},
but for which ALMA observations now reveal the CO distribution in emission \citep{Dent2014}.
Crucially the spatial distribution of this gas was found to exactly match that
of the dust.
The radial distribution is extended over the same range of radii as the mm-sized dust
(roughly 60-130\,au),
and the azimuthal distribution is highly clumped in a manner that matches
that seen in mid-IR images of the distribution of small $\mu$m-sized dust \citep{Telesco2005}.
Since both the $\mu$m-sized dust and CO gas are expected to be short-lived
(the former due to radiation pressure and the latter limited to $\sim 100$\,yr due to
photodissociation from interstellar radiation), the natural conclusion is that both
of these are products of collisions between planetesimals;
i.e., the gas is secondary and continually replenished in the same way as dust in
debris disks.
While the exact mechanism for releasing the gas from icy planetesimals is unknown,
analogy with comets in the Solar System, which can be $\sim 10$\% CO by mass, shows that
the inferred CO production rate of $0.1M_\oplus$/Myr is reasonable as long as
the majority of the CO ice is released as gas as mass passes down the collisional
cascade.

In fact $\beta$ Pic is the only one of the unbiased sample of 6 young A stars
described in \S \ref{ss:dd} to exhibit CO emission, with HR4796A having
a limit of $7M_\oplus$ when translated into molecular H$_2$ \citep{Zuckerman1995,Greaves2000a}.
However, two other young (30-40\,Myr) debris disks, HD21997 and 49 Ceti,
have CO detections at levels slightly above that of $\beta$ Pic (see \S \ref{ss:dd}).
For HD21997 ALMA observations show a distribution of
CO that is not co-located with the mm-sized dust, rather the CO extends
in significantly closer to the star \citep{Kospal2013,Moor2013}.
This suggests that the gas cannot be a product of planetesimal collisions,
and instead must be primordial, which in turn raises the issue of why the
CO has not been photodissociated on short timescales.
In protoplanetary disks, the bulk of the disk material including CO is completely shielded from 
stellar irradiation and therefore the dissociation (and photoevaporation) is gradual,
taking place only in the exposed surface layers of the disk.
In the optically thin regime associated with debris disks, CO photodissociation would be
expected to be more effective unless the CO molecules remain efficiently shielded by
something other than dust.
Whether the optically thin dust phase can be reached fast enough that the gas has not
had time to completely disperse and so is present in sufficient quantities to
effectively shield the CO remains an open question.

There are two other gas species detections amongst debris disks, and both
lie in our young 6 A star sample (\S \ref{ss:dd}):
HD172555 has a detection of OI, and $\eta$ Tel has a
detection of CII \citep{RiviereMarichalar2014}.
The origin of this gas, and in particular whether it is remnant primordial
material or created in collisions, is unclear.
For now we also do not know its radial location, and so whether its presence
is perhaps related to the elevated levels of hot dust seen in both systems.
However, it is a further indication that gas can persist to late times, and
is one of the last things to disappear.

\subsection{Formation of ring-like planetesimal structures}
\label{ss:ring}
Images of the HR4796A debris disk show that the $\mu$m-sized dust
in this 8\,Myr system is concentrated in a narrow ring at $\sim 80$\,au
\citep{Wahhaj2014,Perrin2014}, suggesting that the narrow ring-like structures
(of fractional radial width $\Delta r/r \approx 10$\%)
seen around older main sequence stars like 440\,Myr Fomalhaut \citep{Kalas2005}
were put in place very early on.
However, the observation of a narrowly confined ring of $\mu$m-sized dust does not necessarily
imply that the same is true of the planetesimals from which the dust is produced.
For example, that planetesimal disk could be much broader but contains just one
location where dust is being produced, say the location where Pluto-sized planetesimals recently formed
\citep{Kenyon2002}.
Alternatively dust could be produced throughout a much broader planetesimal disk but
the smallest dust is shepherded into a narrow region through interaction with a residual gas disk
\citep{Takeuchi2001,Lyra2013}.

Broad planetesimal disks (where $\Delta r/r > 1$) do not appear to be the norm
amongst the population, as can be determined from the evolution of 24 and 70~$\mu$m excesses
from main sequence A stars \citep{Rieke2005,Su2006}.
These excesses get fainter exactly as expected if all A stars were born with narrow
rings \citep{Wyatt2007b}, albeit at a different radius and with a different
initial mass for different stars.
This is because all such rings remain at roughly constant brightness until
the largest planetesimals reach collisional equilibrium (which takes a time that depends
on the radius and initial mass of the belt), whereupon the brightness falls off
inversely with age.
Broader planetesimal belts would exhibit a much different behaviour \citep{Kennedy2010},
in particular exhibiting a much slower fall-off in far-IR brightness which is not observed.
Since mm-sized dust is thought to trace the distribution of the planetesimals in a debris disk,
the breadth of the planetesimal belt can be constrained in ALMA observations.
\footnote{This is because in a debris disk the same physics applies to all objects in
the size range from planetesimals down to mm-sized dust, so as long as collisions do not change
significantly the orbits of debris from those of their parent body, then the distribution
of mm-sized dust should be the same as that of the planetesimals.
This is not the case in a protoplanetary disk due to gas forces which result in the
possibility that the distribution of mm-sized dust bears little resemblance to that of the
planetesimals, as discussed in \S \ref{ss:mm}.}
For now ALMA observations of the old narrow dust ring around Fomalhaut show its
planetesimal population is also narrowly confined \citep{Boley2012}, but it has yet
to be demonstrated that the same is true for younger systems for which different processes
could be at play.

The general conclusion that debris disks are narrow does not exclude the
existence of broad disks in some cases, and indeed one
of the young A stars in \S \ref{ss:dd} has a broad disk. 
Even though the disk of $\beta$ Pic is seen edge-on, the ALMA continuum
image can be deprojected using modelling to show that the mm-sized dust in this
system is distributed over a factor of 3-4 in radius
(i.e., $\Delta r/r \approx 1$), with a similar width inferred from the CO velocities
\citep{Dent2014}.
While it is tempting to interpret $\beta$ Pic as having been caught in the act of
evolving toward a narrow ring, the existence of older A stars with broad
disks, such as $\gamma$ Tri at 160\,Myr \citep{Booth2013} and $\kappa$ CrB
at 2.5\,Gyr \citep{Bonsor2013b}, suggests that broad disks are simply another
possible (though on average less common) outcome.
However, it must be cautioned that some of the radial extent observed in debris
disks, particularly at optical to far-IR wavelengths that are sensitive to $\mu$m-sized dust,
may be attributable to radiation forces, and long wavelength observations are required
to ascertain the radial width of the planetesimal belt.

It is worth noting that the existence of broad planetesimal belts does not
preclude that the disks were born as much narrower rings, because the structure
of a debris disk can be strongly influenced by interactions with planets
in the system.
For example, the planet-like object Fomalhaut-b was found to be on a
highly elliptical orbit that takes it across the narrow debris ring
\citep{Kalas2013}.
Dynamical interactions would broaden the disk on a short timescale
that depends on the planet's mass, but
is of order Myr for planets larger than Neptune \citep{Beust2014,Tamayo2014}.
While this is a somewhat extreme configuration, simulations of interactions
of debris disks with highly eccentric planets such as
those which might originate from planet-planet scattering or merging events
show that broad structures are a natural outcome \citep{Pearce2014}.
Moreover, the clump in the $\beta$ Pic disk has been interpreted as evidence
for outward migration of a planet on a circular orbit which swept planetesimals into its
mean motion resonances \citep{Wyatt2003b,Wyatt2006};
that is, the disk's breadth could be a consequence of that migration.
Thus it remains possible that all debris disks are born narrow and
that those we see as broad may just be those in which strong planet interactions
have taken place.
Alternatively planet interactions may aid concentration into a ring, particularly
if there are planets at both the inner and outer edges, and
ring breadth may serve as an indicator of shepherding planet mass \citep{Rodigas2014}.

High resolution imaging of azimuthal structure in young debris disks can be particularly
telling of the processes involved in setting the distribution of planetesimals in
debris disks and can have implications for the processes operating protoplanetary
disks.
For example if the clumpy structure in the $\beta$ Pic disk is confirmed
to originate in resonant trapping of planetesimals during a prior epoch of outward
planet migration, then that migration potentially occurred during the protoplanetary disk phase;
the secular evolution that lead to the warp of that disk must also have had its origins
at an earlier epoch (see \S \ref{ss:phys}).
In this respect it is notable that the clumpy structure in the $\beta$ Pic
disk bears similarities to those of transition disks, since in both
cases the mm-size dust is seen to be concentrated in a clump, with smaller dust
having an axisymmetric distribution \citep{Telesco2005,Casassus2012,vanderMarel2013,Dent2014}.
Given how close these systems are in evolutionary terms it might seem odd
that the two structures have different physical origins, yet that is thought
to be the case.
The large mass of dust observed in the transition disk horseshoes likely precludes
its origin in resonantly trapped planetesimals, and the low gas mass in $\beta$ Pic
likely precludes its clump having an origin in pressure-induced trapping.

The timing and mechanism for the planetesimals to become concentrated
in a ring (or for planetesimals to become depleted at other locations)
is not well constrained, as discussed in \S \ref{ss:phys} and \S \ref{ss:mm}.
For example, some planetesimals are likely to have been present in radially confined
concentrations very early on in the protoplanetary disk phase, perhaps
at the sites of snow-lines or pressure bumps
\citep{Ros2013,Drazkowska2013}.
Given the large amount of dust present in a protoplanetary disk, any planetesimal
structures would be embedded in this and so only be revealed once the
bulk of the dust had dispersed. 
However, it remains plausible that a significant population of planetesimals
is created during dispersal of the protoplanetary disk (see \S \ref{ss:mm}).

\section{Summary}
\label{s:summ}
This paper discussed the stages that occur as protoplanetary disks
transition to debris disks.
The picture that emerges has several caveats, notably with respect to the
order of the stages and whether they take place consecutively or concurrently,
but it is worth presenting a strawman evolutionary sequence to use as the
basis for further discussion.

It is likely that both planetesimals and planet-sized objects are already present
before the onset of the dispersal of the protoplanetary disk.
This does not mean that they are necessarily present, just that if they form at all
then this probably occurs before the transition.
The existence of planetesimals is difficult to confirm, because they are masked by
the large quantities of small dust that is also present, but we noted that
these could be stirred by any planets present, and so that some of the small dust observed
in this phase could arise from those planetesimals.
The planets themselves may be easier to confirm,
because their gravitational influence can leave discernable
signatures on the disk structure \citep[e.g.,][]{Wolf2005}, but only if
they are sufficiently massive.

The first stage is then the depletion of material from the inner regions
of the disk.
Accretion onto the star is ongoing in the transition disk phase, suggesting that these inner
regions still contain gas, but are absent of small dust.
Probably this occurs when the planetary system has grown sufficiently
to curtail the replenishment of small dust in the inner regions, and to
disrupt the inward flow of small grains from the outer disk through its effect
on the gas disk structure \citep{Pinilla2013,Owen2014},
though this explanation is not without problems \citep{Clarke2013}.

The second stage is the depletion of the mm-sized dust from the outer disk.
Whether this mass ends up in large objects (i.e., planetesimals or planets) or
ground down into small dust is unknown, though this loss likely occurs
quickly and from a mass budget perspective it is plausible that the majority
of the mass ends up as planetesimals (see \S \ref{ss:mm}).
The concentration of mm-sized dust in gas structures in transition disks
could aid both scenarios \citep{Birnstiel2013,Zhu2014}.

Dust in the inner regions of systems beyond the transition disk phase is thought
to be produced by the break-up of planetesimals and perhaps even planets,
rather than a remnant of the protoplanetary dust.
That is, moderate levels of hot dust can persist onto the main sequence, but
this is a debris population.
Hot dust with a debris-like origin can also be present in protoplanetary disks,
and studying such dust at all evolutionary phases may provide a window into the
possible presence and formation mechanism of small rocky
planets close to the star, the statistics for which remain sparse for A stars
\citep{Lagrange2009,Johnson2011,Mulders2014}.

One stage which is particularly unclear in its timing is the concentration of
the planetesimals into a narrow ring, which is the most common outcome
of the evolution (as opposed to a broad disk).
This structure could have been imprinted early on in the protoplanetary disk
phase, as appears to be the case in HL Tau, or fixed during the dispersal of the gas disk.

The last (and least understood) stage is the removal of the gas.
The prevalence of gas around young A star debris disks shows that this component
persists beyond the protoplanetary disk phase.
However, most of this gas is thought to be secondary (i.e., released from
its storage as ices in solid planetesimals) and so a debris-related phenomenon. 
Nevertheless, the persistence of gas around HD21997 which is probably primordial
suggests that this is the last component to disperse.

We proposed that the distinction between protoplanetary and debris disks
is primarily a question of the existence of large quantities of primordial gas,
where we define large as being sufficient to entrain small dust grains in the disk
and so damping collision velocities maintaining it at elevated levels.
That is, the level of stirring in the disk is a consequence of the nature
of the disk, but not its defining property.

Since the mass of the gas disk is difficult to measure the observational classification
must rely on other tracers.
We defined an empirical classification using flux ratios at 12 and 70~$\mu$m, with
debris disks requiring $R_{12}<3$ and $R_{70}<2000$ (noting that this
definition only applies to A stars).
However, we also cautioned that interpretation of SEDs may be degenerate
if the SED of a protoplanetary disk with an inner hole and a
reasonable degree of dust settling is indistinguishable from that of a debris disk.
Detailed study of the gas content in such systems would be needed to ascertain
their evolutionary status.

%
\acknowledgments
The authors are grateful for support from the European Union through ERC grant number 279973.


\bibliographystyle{spr-mp-nameyear-cnd}  
\bibliography{lib}                

\end{document}